\documentclass[preprint,superscriptaddress,amsmath,amssymb,aps,prb]{revtex4-2}

\usepackage{graphicx}
\usepackage[version=4]{mhchem}
\usepackage[dvipsnames]{xcolor}
\usepackage{soul}

\definecolor{dartmouthgreen}{rgb}{0.05, 0.5, 0.06}

\begin{document}

\title{Simulating charging characteristics of lithium iron phosphate by \\ electro-ionic optimization on a quantum annealer}

\author{Tobias Binninger}
\email{t.binninger@fz-juelich.de}
\affiliation{Theory and Computation of Energy Materials (IET-3), Institute of Energy Technologies, Forschungszentrum J\"ulich GmbH, 52425 J\"ulich, Germany}
\affiliation{J\"ulich Aachen Research Alliance JARA Energy
\& Center for Simulation and Data Science (CSD), 52425 J\"ulich, Germany}

\author{Yin-Ying Ting}
\affiliation{Theory and Computation of Energy Materials (IET-3), Institute of Energy Technologies, Forschungszentrum J\"ulich GmbH, 52425 J\"ulich, Germany}
\affiliation{J\"ulich Aachen Research Alliance JARA Energy
\& Center for Simulation and Data Science (CSD), 52425 J\"ulich, Germany}

\author{Konstantin K\"oster}
\affiliation{Materials Synthesis and Processing (IMD-2), Institute of Energy Materials and Devices, Forschungszentrum J\"ulich GmbH, 52425 J\"ulich, Germany}
\affiliation{MESA+ Institute, University of Twente, Hallenweg 15, 7522, NH, Enschede, The Netherlands}

\author{Nils Bruch}
\affiliation{Theory and Computation of Energy Materials (IET-3), Institute of Energy Technologies, Forschungszentrum J\"ulich GmbH, 52425 J\"ulich, Germany}
\affiliation{J\"ulich Aachen Research Alliance JARA Energy
\& Center for Simulation and Data Science (CSD), 52425 J\"ulich, Germany}

\author{Payam Kaghazchi}
\affiliation{Materials Synthesis and Processing (IMD-2), Institute of Energy Materials and Devices, Forschungszentrum J\"ulich GmbH, 52425 J\"ulich, Germany}
\affiliation{MESA+ Institute, University of Twente, Hallenweg 15, 7522, NH, Enschede, The Netherlands}

\author{Piotr M. Kowalski}
\affiliation{Theory and Computation of Energy Materials (IET-3), Institute of Energy Technologies, Forschungszentrum J\"ulich GmbH, 52425 J\"ulich, Germany}
\affiliation{J\"ulich Aachen Research Alliance JARA Energy
\& Center for Simulation and Data Science (CSD), 52425 J\"ulich, Germany}

\author{Michael H. Eikerling}
\affiliation{Theory and Computation of Energy Materials (IET-3), Institute of Energy Technologies, Forschungszentrum J\"ulich GmbH, 52425 J\"ulich, Germany}
\affiliation{J\"ulich Aachen Research Alliance JARA Energy
\& Center for Simulation and Data Science (CSD), 52425 J\"ulich, Germany}
\affiliation{Chair of Theory and Computation of Energy Materials, Faculty of Georesources and Materials Engineering, RWTH Aachen University, Intzestrasse 5, 52072 Aachen, Germany}

\begin{abstract}
The rapid evolution of quantum computing hardware opens up new avenues in the simulation of energy materials. Today's quantum annealers are able to tackle complex combinatorial optimization problems. A formidable challenge of this type is posed by materials with site-occupational disorder for which atomic arrangements with a low, or lowest, energy must be found. In this article, a method is presented for the identification of the correlated ground-state distribution of both lithium ions and redox electrons in lithium iron phosphate (LFP), a widely employed cathode material in lithium-ion batteries. The point-charge Coulomb energy model employed correctly reproduces the LFP charging characteristics. As is shown, grand-canonical transformation of the energy cost function makes the combinatorial distribution problem solvable on quantum annealing (QA) hardware. The QA output statistics follow a pseudo-thermal behavior characterized by a problem-dependent effective sampling temperature, which has bearings on the estimated scaling of the QA performance with system size. This work demonstrates the potential of quantum computation for the joint optimization of electronic and ionic degrees of freedom in energy materials.
\end{abstract}

\maketitle

\section*{Introduction}

\begin{figure*}[t]
\centering
\includegraphics[width=0.85\textwidth]{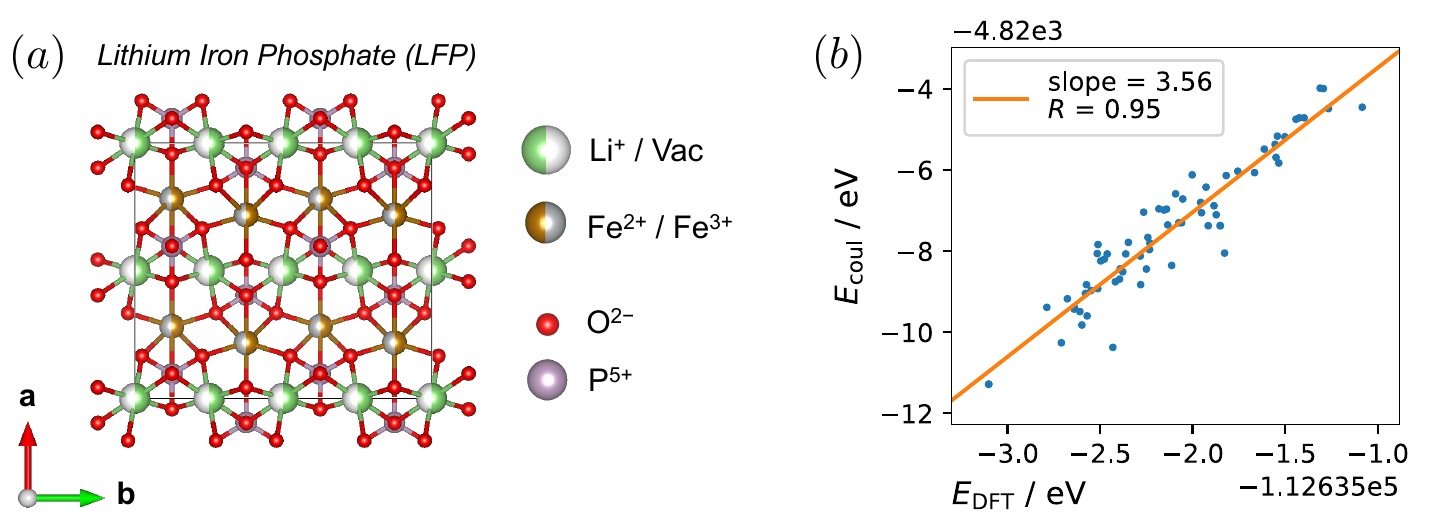}
\caption{(a) Simulation cell of 
\ce{Li_{x}FePO4} ($1\times2\times2$ supercell) comprising a total of 16 Li sites (either occupied or vacant) and 16 Fe sites (occupied by either \ce{Fe(II)} or \ce{Fe(III)}). The formal ionic charges used in the Coulomb energy model are indicated. (b) Comparison of Coulomb ($E_{\mathrm{coul}}$) \textit{vs.} DFT ($E_{\mathrm{DFT}}$) energies for different configurations of Li ions in the semi-lithiated $1\times4\times1$ supercell. For the Coulomb energies, the minimum energy distributions of the redox electrons across the iron sites for the given (frozen) Li configurations were determined by full enumeration.}
\label{fig_LFP_model}
\end{figure*}

With the fast-paced development of quantum computing hardware, methods and algorithms for leveraging quantum computation (QC) in materials research are gaining attention. The exploration of vast configuration spaces  in elemental composition and atomic structure is an overarching challenge in materials optimization~\cite{daviesComputationalScreeningAll2016a, gusevOptimalityGuaranteesCrystal2023} that is particularly well suited for QC approaches. Such problems are commonly addressed by discretizing the configuration space in terms of atomic \textit{sites}, which can be occupied by different atom species. As an example, Fig.~\ref{fig_LFP_model}a shows a simulation cell of lithium iron phosphate (LFP) comprising a number of lithium ion sites (indicated by green-white spheres) that can be either occupied or vacant, depending on the state of charge, \textit{i.e.}, the lithium stoichiometry, of the material. For each site and species, a binary variable $x_i\in\left\{0,1\right\}$ represents whether the given site is occupied ($x_i = 1$) by said species, or not ($x_i = 0$). The physical quantity of interest, often the total energy, $E$, is then expressed as a function of the site occupation variables, $E(\{x_i\})$, written in the form of a cluster expansion (CE)~\cite{connollyDensityfunctionalTheoryApplied1983, sanchezGeneralizedClusterDescription1984, ferreiraFirstprinciplesCalculationAlloy1989, laksEfficientClusterExpansion1992},
\begin{align}
E(\{x_i\})\, =\, J_0 + \sum_{i} \,J_{i}\,x_i + \sum_{i<j} \,J_{ij}\,x_i\,x_j + \sum_{i<j<k} \,J_{ijk}\,x_i\,x_j\,x_k + \cdots \ ,
\label{eq_cluster_expansion}
\end{align}
which is the analogue of a Taylor series expansion for functions of discrete-valued variables. In practice, the expansion coefficients $J_{\alpha}$, termed effective cluster interactions (ECI)~\cite{tepeschModelConfigurationalThermodynamics1995}, are fitted to best match the \textit{ab initio} energies, \textit{e.g.}, from density functional theory (DFT) calculations, of a certain sub-set of configurations. The cluster expansion model, Eq.~\eqref{eq_cluster_expansion}, then enables the use of classical heuristics, such as Monte Carlo algorithms, for the search of the lowest-energy configuration (ground state), or the thermodynamical sampling of the system's free energy~\cite{zhouConfigurationalElectronicEntropy2006}. This approach has been widely used for the description of materials with substitutional disorder, such as metallic alloys~\cite{ferreiraFirstprinciplesCalculationAlloy1989, vandewalleAlloyTheoreticAutomated2002, blumMixedbasisClusterExpansion2004, sanchezClusterExpansionConfigurational2010} or cathode materials for Li-ion batteries~\cite{urbanComputationalUnderstandingLiion2016, leeRapidlyConvergentCluster2017} that comprise a huge number of possible atomic (or ionic) arrangements across available lattice sites. It is worth noting, however, that the CE method is of more general applicability in materials optimization; CE approaches have been used, \textit{e.g.}, to effectively search for new catalyst materials with optimized compositions in a multi-elements chemical space~\cite{choubisaAcceleratedChemicalSpace2023}.

Quantum annealing (QA), a form of adiabatic quantum computation (AQC)~\cite{albashAdiabaticQuantumComputation2018}, is a promising alternative to classical heuristic methods for solving discrete optimization problems. In QA, the quantum system of qubits is initially prepared in the ground state $|\mathcal{G}_0\rangle$ of some simple initial Hamiltonian $\mathcal{H}_0$, which is then tuned with a certain schedule $s(t)$ (where $s(0)=0$ and $s(t_{\mathrm{ann}})=1$) to the target Hamiltonian $\mathcal{H}_1$ that encodes the problem at hand. According to the adiabatic theorem of quantum mechanics, the system is expected to remain in the instantaneous ground state of the time-dependent Hamiltonian $\mathcal{H}(t) = (1-s(t))\,\mathcal{H}_0 + s(t)\,\mathcal{H}_1$ if the tuning is sufficiently slow. In particular, at the end of the schedule, the qubit system would be found in the ground state of the target Hamiltonian, $|\Psi(T)\rangle = |\mathcal{G}_1\rangle$~\cite{albashAdiabaticQuantumComputation2018}. Current QA devices, such as the quantum annealers by D-Wave Systems Inc., are specifically designed for solving (classical) quadratic unconstrained binary optimization (QUBO) problems that are represented by a quadratic cost function of a set of classical binary variables,
\begin{align}
E(\{x_i\})\, =\, E_0 + \sum_{i} \,Q_{i}\,x_i + \sum_{i<j} \,Q_{ij}\,x_i\,x_j \ .
\label{eq_qubo_cost_function}
\end{align}
It is worth noting that the QUBO form corresponds to a second-order (approximated) cluster expansion, \textit{cf.} Eq.~\eqref{eq_cluster_expansion}. Omitting the unimportant constant, $E_0$, the QUBO cost function~\eqref{eq_qubo_cost_function} is mapped to a Hamiltonian
\begin{align}
\mathcal{H} \ =\ \sum_{i} Q_{i}\,\hat{\sigma}_z^{(i)} + \sum_{i<j} Q_{ij}\,\hat{\sigma}_z^{(i)}\hat{\sigma}_z^{(j)}
\label{eq_qubo_Hamiltonian}
\end{align}
comprising only $\hat{\sigma}_z$ qubit operators, for which reason $\mathcal{H}$ commutes with all $\hat{\sigma}_z^{(i)}$ that define the computational basis. The respective ground state (as well as all eigenstates) is therefore a simple product state (\textit{e.g.}, $|\mathcal{G}_1\rangle = |1,0,1,1,\dots\rangle$), which can be interpreted as a classical bit sequence representing the minimum of the QUBO cost function~\eqref{eq_qubo_cost_function}. In practice, imperfections and noise produce statistics in the output that render QA a heuristic method~\cite{benedettiEstimationEffectiveTemperatures2016, benedettiQuantumAssistedLearningHardwareEmbedded2017, bruggerOutputStatisticsQuantum2022}.

Examples for the use of QA and QA-inspired approaches to tackle combinatorial problems in materials research include, \textit{e.g.}, protein folding~\cite{perdomo-ortizFindingLowenergyConformations2012, robertResourceefficientQuantumAlgorithm2021}, conformational sampling in mixtures of polymer chains~\cite{michelettiPolymerPhysicsQuantum2021}, the design of metamaterials with optimized thermal emission/absorption spectra~\cite{kitaiDesigningMetamaterialsQuantum2020}, chemical space search of organic molecular compounds~\cite{hatakeyama-satoTacklingChallengeHuge2021}, crystal structure prediction~\cite{gusevOptimalityGuaranteesCrystal2023, couzinieAnnealingPredictionGrand2024}, and the compositional optimization of mixed-metal oxide catalysts~\cite{choubisaAcceleratedChemicalSpace2023}. Likewise, QA methods have been developed for the optimization of vacancy distributions in graphene~\cite{carnevaliVacanciesGrapheneApplication2020, caminoQuantumComputingMaterials2023}, metal atom arrangements in alloys~\cite{ichikawaAcceleratingOptimalElemental2024}, and ionic configurations in cathode materials for Li-ion batteries~\cite{binningerOptimizationIonicConfigurations2024}. The treatment of constraints represents a major challenge in QA, which can only be implemented in a ``soft'' manner by adding suitable penalty terms to the QUBO cost function. A target number of particles, or target stoichiometry, is a frequently encountered constraint in materials research. For a set of site occupation variables, $x_i\in\left\{0,1\right\}$, the number of particles is simply given by the total number of occupied sites, \textit{i.e.},
\begin{align}
N = \sum_{i} \,x_i \ ,
\label{eq_Hamming_weight}
\end{align}
which, more generally, is known as the Hamming weight of the bit string. To achieve a certain target value, $N^{\mathrm{target}}$, a quadratic penalty term is added to the cost function,
\begin{align}
E_{\mathrm{penalty}} = \lambda\,\left( \sum_{i} \,x_i - N^{\mathrm{target}} \right)^2 \ ,
\label{eq_energy_penalty}
\end{align}
which increases the cost of solutions that violate the constraint, whilst leaving the target subspace unchanged. Here, $\lambda$ is a parameter controlling the strength of the constraint. A major disadvantage of this approach results from the quadratic terms $\sum_{i<j} 2\lambda\,x_i x_j$ in Eq.~\eqref{eq_energy_penalty}, which add a value of $2\lambda$ to all off-diagonal elements $Q_{ij}$ (with $i<j$) of the QUBO matrix. If the required value of $\lambda$ is much larger than the magnitude of the $Q_{ij}$, the penalty term can mask the basic cost term~\eqref{eq_qubo_cost_function} and render the optimization by QA ineffective. To mitigate this problem, some of the authors recently introduced a (linear) Legendre transformation,
\begin{align}
E(\{x_i\}) - \mu \sum_{i} \,x_i \ ,
\label{eq_grand_canonical_cost_function}
\end{align}
yielding a grand-canonical cost function, $E-\mu N$, with a chemical potential $\mu$~\cite{binningerOptimizationIonicConfigurations2024}. Choosing the latter equal to the local slope of the $E$ \textit{vs.} $N$ curve, $\mu = \partial E/\partial N$, the grand-canonical cost function~\eqref{eq_grand_canonical_cost_function} becomes flat around $N^{\mathrm{target}}$. The penalty term~\eqref{eq_energy_penalty} with a minor value of $\lambda$ is then sufficient to produce a minimum at the target particle number in the overall cost function,
\begin{align}
E_{\mathrm{tot}}(\{x_i\}) = E(\{x_i\}) - \mu \sum_{i} \,x_i + \lambda\,\left( \sum_{i} \,x_i - N^{\mathrm{target}} \right)^2 \ .
\label{eq_total_cost_function}
\end{align}
Since the Legendre transformation~\eqref{eq_grand_canonical_cost_function} only depends on the total particle number, it corresponds to a constant shift within the target subspace of $\sum_{i} \,x_i = N^{\mathrm{target}}$ and thus does not interfere with the basic QUBO. In said previous work~\cite{binningerOptimizationIonicConfigurations2024}, the arrangement of Li ions in lithium cobalt oxide (LCO) was sampled using a D-Wave quantum annealer. A simple Coulomb interaction model with formal ionic charges was used to describe the energy of different configurations. The Legendre transformation of the cost function, Eq.~\eqref{eq_grand_canonical_cost_function}, enabled the successful identification of the ionic ground-state configuration for the target stoichiometry of $\ce{Li_{1/2}CoO2}$ among a total number of $2^{36} \approx 7\times10^{10}$ configurations in a simulation cell with 36 Li sites. Despite the simplicity of the Coulomb model, the respective ionic ground state was found to agree with the alternating row-like Li ordering known from experiment. 

For the previously studied case of LCO, only ionic degrees of freedom were optimized, while the electronic charge was equally distributed among the redox-active cobalt ions with an effective average charge of $+3.5\,\mathrm{e}$ for the semi-lithiated state of LCO (\ce{Li_{1/2}CoO2})~\cite{binningerOptimizationIonicConfigurations2024}. This treatment was justified by the metallic character of \ce{Li_{x}CoO2} for $x < 0.75$~\cite{marianettiFirstorderMottTransition2004} and the corresponding delocalization of the electronic charge at the Fermi level. In the present work, the grand-canonical QA approach is extended to the simulation of lithium iron phosphate (LFP), another standard cathode material for Li-ion batteries. In contrast to LCO, the charging/discharging of LFP occurs via a direct phase transition between \ce{LiFePO4} (triphylite, with formal \ce{Fe^{2+}} valency) and \ce{FePO4} (heterosite, with formal \ce{Fe^{3+}} valency)~\cite{padhiPhosphoolivinesPositiveElectrodeMaterials1997}, due to a wide miscibility gap at room temperature~\cite{yamadaRoomtemperatureMiscibilityGap2006, kowalskiElectrodeElectrolyteMaterials2021}. At higher temperatures, the existence of \ce{Li_{x}FePO4} solid solutions was demonstrated~\cite{delacourtExistenceTemperaturedrivenSolid2005, kowalskiElectrodeElectrolyteMaterials2021}. Electronic disproportionation results in the simultaneous presence of \ce{Fe^{2+}} and \ce{Fe^{3+}} species in mixed-valence \ce{Li_{x}FePO4}. Zhou et al.~\cite{zhouConfigurationalElectronicEntropy2006} included the localized electronic degrees of freedom, \textit{i.e.}, the distribution of redox electrons (\ce{Fe^{2+}}) and holes (\ce{Fe^{3+}}) in a cluster expansion energy model parametrized by DFT$+U$ calculations. They showed that the \ce{LiFePO4}/\ce{FePO4} phase separation at low temperature results from the attractive interaction between \ce{Li^+} and \ce{e^-}, making it energetically favourable for lithium ions to accumulate together with the redox electrons (\ce{Fe^{2+}}) in a \ce{LiFePO4} phase, which is consistent with phase separation. At higher temperature, however, the combined electronic and ionic configurational entropy stabilizes the solid-solution phase of \ce{Li_{x}FePO4}. 

Herein, quantum annealing is used to simulate the charging characteristics of LFP in a joint electro-ionic combinatorial optimization, employing a point-charge Coulomb energy model. Performing two Legendre transformations in charged and charge-neutral variations of \ce{Li^+} and \ce{e^-} particle numbers, the problem is made feasible for the identification of the configurational ground state on a D-Wave quantum annealer. This work takes a step towards leveraging quantum computing methods for the joint treatment of electronic and ionic degrees of freedom in redox-active battery materials.

\section*{Results}

In this study, a classical point-charge Coulomb model is used to describe the energy of different distributions of Li ions and redox electrons across available Li and Fe sites, respectively, in the LFP simulation cell, as described in detail in the Methods section. Albeit in a highly simplified fashion, the point-charge Coulomb model is sufficient for jointly treating electronic and ionic degrees of freedom in the simulation of battery materials at different state of charge (SOC) with the QA method.

\subsection*{Validation of the LFP Coulomb model}

Before shifting the focus to the QA results, the Coulomb energy model is validated by comparison against DFT calculations. The model is then shown to correctly describe the known two-phase charging characteristics of LFP, and, thus, to capture the essential physics of LFP charging/discharging processes.

\subsubsection*{Comparison between Coulomb and DFT energies}
Fig.~\ref{fig_LFP_model}b presents a plot of point-charge Coulomb energies \textit{vs.} DFT energies for 68 different arrangements of Li ions in the semi-lithiated $1\times4\times1$ supercell. For the Coulomb energies, the respective minimum energy distributions of the compensating redox electrons across the iron sites were determined by full enumeration of all possible electronic configurations. As observed, the energies from the electro-ionic Coulomb model are strongly correlated to the DFT energies, as indicated by a Pearson correlation coefficient of $R = 0.95$. This means that configurations with low Coulomb energy are also likely to have low DFT energy. In particular, among the set of Li-ion configurations considered in this comparison, the configuration with the minimum Coulomb energy, corresponding to the global ground state of the Coulomb model, also had the lowest DFT energy. The electro-ionic Coulomb ground state is, thus, a likely candidate for the Li-ion arrangement with minimum, or at least close to minimum, DFT energy. In absolute terms, however, the variations in Coulomb energies among different configurations are much larger than the respective variations in DFT energies, which is expected due to the lack of dielectric screening in the Coulomb model. The slope in Fig.~\ref{fig_LFP_model}b can thus be interpreted as an effective dielectric constant of about $\epsilon_r = 3.56$ caused by electronic orbital relaxation, which is implicitly included in DFT energies. It should be noted, however, that materials such as LFP that contain multi-valent ions of the same element (the iron redox centers in LFP) are notoriously challenging for DFT simulations, not only because of the strong electronic correlation effects~\cite{kowalskiElectrodeElectrolyteMaterials2021}, but also the electronic disproportionation that results in a combinatorial optimization problem for the distribution of the redox electrons in a similar way as for the point-charge Coulomb model. While the present work investigates QA approaches for Coulomb optimization, similar QA methods should be developed for DFT minimization problems in the future.

\begin{figure*}[t]
\centering
\includegraphics[width=0.85\textwidth]{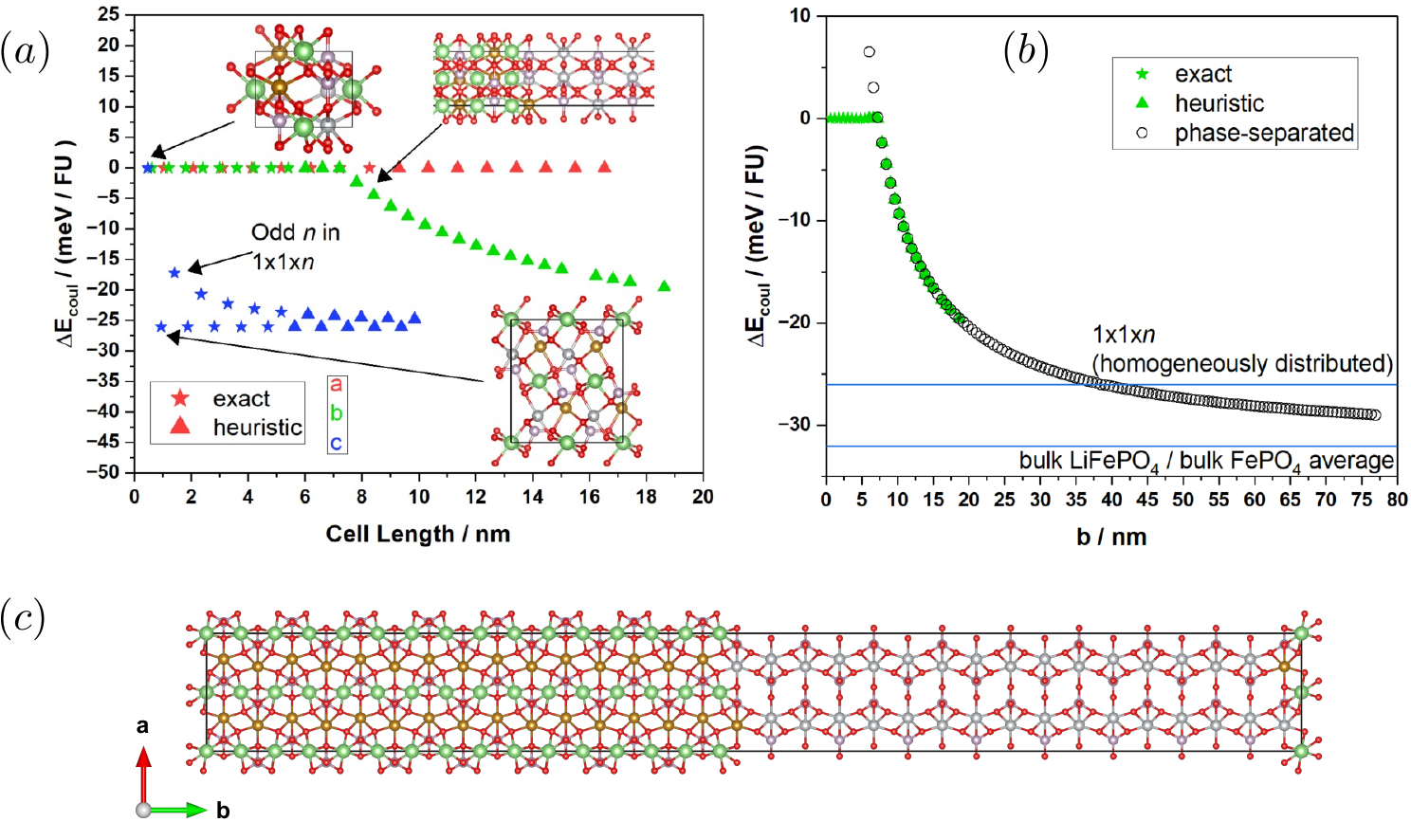}
\caption{(a) Coulomb energies of minimum energy configurations for semi-lithiated LFP (\ce{Li_{0.5}FePO4}) as a function of the size (length) of the simulation cell along different lattice directions. The energies are given per \ce{Li_{0.5}FePO4} formula unit (FU) on a relative scale using the ground state configuration for the $1 \times 1 \times 1$ unit cell as a reference. The simulation cells had elongated shapes of dimensions $n \times 1 \times 1$, $1 \times n \times 1$, and $1 \times 1 \times n$ with increasing multiplicities ($n$) along the $a$ (red data points), $b$ (green data points) and $c$ (blue data points) directions, respectively. The respective minimum energy configurations are shown. (b) Extended plot of the minimum energy as a function of cell size in $b$ direction. The energy of the homogeneously distributed ground-state configuration for the $1 \times 1 \times n$ cells (with even $n$) and the average energy of bulk \ce{LiFePO4} and bulk \ce{FePO4} are given as reference lines. (c) Minimum energy configuration for a $1 \times 16 \times 1$ cell with a phase separation into \ce{LiFePO4} and \ce{FePO4} domains. Color labelling for different ion species as in Fig.~\ref{fig_LFP_model}a.}
\label{fig_LFP_ground_state}
\end{figure*}

\subsubsection*{Two-phase charging characteristics}
Fig.~\ref{fig_LFP_ground_state}a shows the Coulomb energies, normalized per LFP formula unit (FU), of minimum energy configurations for semi-lithiated LFP as a function of the length of the simulation cell along different lattice directions. Ground state configurations were determined by classical optimization with the Gurobi software~\cite{GurobiOptimizerReference2024}. For smaller cell sizes, solutions obtained were proven to represent the exact optimum (star data points in Fig.~\ref{fig_LFP_ground_state}a). It is observed that minimum energy configurations are strongly dependent on size and shape of the simulation cell. For $n \times 1 \times 1$ cells elongated in $a$ direction (red data points), the ground state configuration of the $1 \times 1 \times 1$ unit cell (shown as an inset in Fig.~\ref{fig_LFP_ground_state}a) was preserved and simply repeated $n$-times, resulting in a constant energy per FU. For cells extended in $c$ direction ($1 \times 1 \times n$, blue data points), a different behavior is observed between cells with even and odd values of $n$. For even $n$, the ground state is a homogeneously distributed configuration independent of $n$ (shown as an inset in Fig.~\ref{fig_LFP_ground_state}a), at least up to a $c$-length of about $10\,\mathrm{nm}$, with a respective per-FU energy significantly lower than for the $n \times 1 \times 1$ ground state. For odd $n$, the ground state configuration is $n$-dependent with an energy decreasing from the initial value of the $1 \times 1 \times 1$ unit cell towards the energy of the constant ground state for even $n$.

For cells elongated in $b$ direction ($1 \times n \times 1$, green data points), the minimum energy configuration of the $1 \times 1 \times 1$ unit cell is maintained up to a critical multiplicity of $n=12$ (length of $7.212\,\mathrm{nm}$). At larger cell dimensions, $n>12$, a phase-separated ground state emerges with one \ce{LiFePO4} and one \ce{FePO4} domain separated along the $b$ direction, which is shown in Fig.~\ref{fig_LFP_ground_state}c for the $1 \times 16 \times 1$ cell. The respective energy decreases with increasing $n$, as shown on an extended range in Fig.~\ref{fig_LFP_ground_state}b, and eventually converges towards the average energy of bulk \ce{LiFePO4} and bulk \ce{FePO4} (bottom horizontal reference line in Fig.~\ref{fig_LFP_ground_state}b). This behavior can be explained by decomposing the energy of the phase-separated cell as~\cite{abdellahiParticlesizeMorphologyDependence2014}
\begin{align}
E_{\mathrm{cell}} = \frac{N_{\mathrm{f}}}{2} E_{\ce{LiFePO4}} + \frac{N_{\mathrm{f}}}{2} E_{\ce{FePO4}} + 2A\,\gamma \ ,
\end{align}
where $E_{\ce{LiFePO4}}$ and $E_{\ce{FePO4}}$ are the bulk energies per formula unit of \ce{LiFePO4} and \ce{FePO4}, respectively, $N_{\mathrm{f}}$ is the overall number of \ce{Li_{0.5}FePO4} formula units in the simulation cell, and $\gamma$ is the interfacial energy per area ($A$) between the \ce{LiFePO4} and \ce{FePO4} domains, where the factor of 2 accounts for the two interfaces per cell. With increasing $N_{\mathrm{f}}$, \textit{i.e.}, cell length, the interface contribution to $E_{\mathrm{cell}}$ decreases, and the energy normalized by $N_{\mathrm{f}}$ converges towards the average of the bulk energies, $(E_{\ce{LiFePO4}} + E_{\ce{FePO4}})/2 = -301.987\,\mathrm{eV}$, with $E_{\ce{LiFePO4}} = -294.067\,\mathrm{eV}$ and $E_{\ce{FePO4}} = -309.907\,\mathrm{eV}$. Using $E_{\mathrm{cell}}/N_{\mathrm{f}} = -301.963\,\mathrm{eV}$ for the phase-separated configuration in a $1 \times 16 \times 1$ cell ($N_{\mathrm{f}} = 64$) with an $ac$ interfacial area of $A = 0.485\,\mathrm{nm^2}$, the interfacial energy is estimated as $\gamma_{ac} = 1.584\,\mathrm{eV/nm^2} = 254\,\mathrm{mJ/m^2}$. Beyond a cell length of about $40\,\mathrm{nm}$ in $b$ direction, the per-atom energy of the phase-separated configuration drops below the energy of the homogeneously distributed $1 \times 1 \times n(\mathrm{even})$ ground state, indicated as a horizontal line in Fig.~\ref{fig_LFP_ground_state}b. The Coulomb model thus predicts a critical particle size of about $40\,\mathrm{nm}$ above which LFP at a half-lithiated SOC decomposes into domains of \ce{LiFePO4} and \ce{FePO4}. For smaller LFP particles, the semi-lithiated solid solution is predicted to be stable even at low temperatures.

These results are in good agreement with the known two-phase charging characteristics of LFP, exhibiting a direct transition between the \ce{LiFePO4} and \ce{FePO4} end members~\cite{padhiPhosphoolivinesPositiveElectrodeMaterials1997, yamadaRoomtemperatureMiscibilityGap2006, zhouConfigurationalElectronicEntropy2006}. Within the Coulomb model, this is reflected by the fact that the phase-separated configuration becomes lowest in energy for large simulation cells and approaches the absolute minimum defined by the bulk average of \ce{LiFePO4} and \ce{FePO4}. As already pointed out in previous studies~\cite{zhouConfigurationalElectronicEntropy2006, wangDecisiveRoleElectrostatic2023}, this is explained by the attractive interaction between \ce{Li^+} and \ce{e^-}, driving the condensation of lithium ions and redox electrons into \ce{LiFePO4} domains within the \ce{FePO4} host structure. The Coulomb model also correctly predicts the preferred $b$ direction of the phase separation with a domain interface in the $ac$ plane, in agreement with previous theoretical studies~\cite{cogswellCoherencyStrainKinetics2012, abdellahiParticlesizeMorphologyDependence2014, wangDecisiveRoleElectrostatic2023}. Surprisingly, even the predicted critical particle size of about $40\,\mathrm{nm}$, below which a solid-solution phase is formed, is in very good agreement with experimental observations, \textit{e.g.}, by Gibot \textit{et al.}~\cite{gibotRoomtemperatureSinglephaseLi2008} and Ichitsubo \textit{et al.}~\cite{ichitsuboWhatDeterminesCritical2013}, who reported single-phase charging/discharging behavior for LFP nanoparticles of about $40\,\mathrm{nm}$ and $10\,\mathrm{nm}$ in size, respectively. On the other hand, based on DFT calculations, Abdellahi \textit{et al.}~\cite{abdellahiParticlesizeMorphologyDependence2014} reported an extremely small value of the interfacial energy $\gamma_{ac} = 7\,\mathrm{mJ/m^2}$, which would yield a much lower value of the critical particle size than experimentally observed. It was therefore concluded that the experimental results could only be explained by accounting for the additional effect of coherency strain at the two-phase boundary~\cite{abdellahiParticlesizeMorphologyDependence2014, abdellahiThermodynamicStabilityIntermediate2016} resulting from the difference in volumes of the two phases and the thermodynamics of solid solution~\cite{kowalskiElectrodeElectrolyteMaterials2021}. To compare their DFT-based value for $\gamma_{ac}$ to the one obtained here from the Coulomb model, \textit{viz.} $\gamma_{ac} = 254\,\mathrm{mJ/m^2}$, the latter has to be rescaled with the effective dielectric constant, $\epsilon_r = 3.56$, as obtained from Fig.~\ref{fig_LFP_model}b, yielding a value of $\gamma_{ac}/\epsilon_r = 71\,\mathrm{mJ/m^2}$. This value is larger by a factor of 10 than the value reported by Abdellahi \textit{et al.}~\cite{abdellahiParticlesizeMorphologyDependence2014}. It is important to note, however, that the value of $71\,\mathrm{mJ/m^2}$ estimated here is more reasonable when compared to the reported values for the $bc$ and $ab$ interfaces ($115\,\mathrm{mJ/m^2}$ and $95\,\mathrm{mJ/m^2}$, respectively) from the same study~\cite{abdellahiParticlesizeMorphologyDependence2014}. Based on the present results, it is thus suggested to reconsider, whether interfacial electric/chemical energy or coherency strain effects are the dominating factors behind the single-phase behavior of nano-sized LFP particles.

To sum up the findings of this section, while the LFP charging characteristics have been accurately described before in numerous DFT studies, it is remarkable that the simple point-charge Coulomb model correctly predicts the main features of (i) bulk \ce{LiFePO4}/\ce{FePO4} phase separation, (ii) a preferred phase interface in the $ac$ plane, and (iii) a solid-solution behavior for particles below about $40\,\mathrm{nm}$.

\begin{figure*}[t]
\centering
\includegraphics[width=0.95\textwidth]{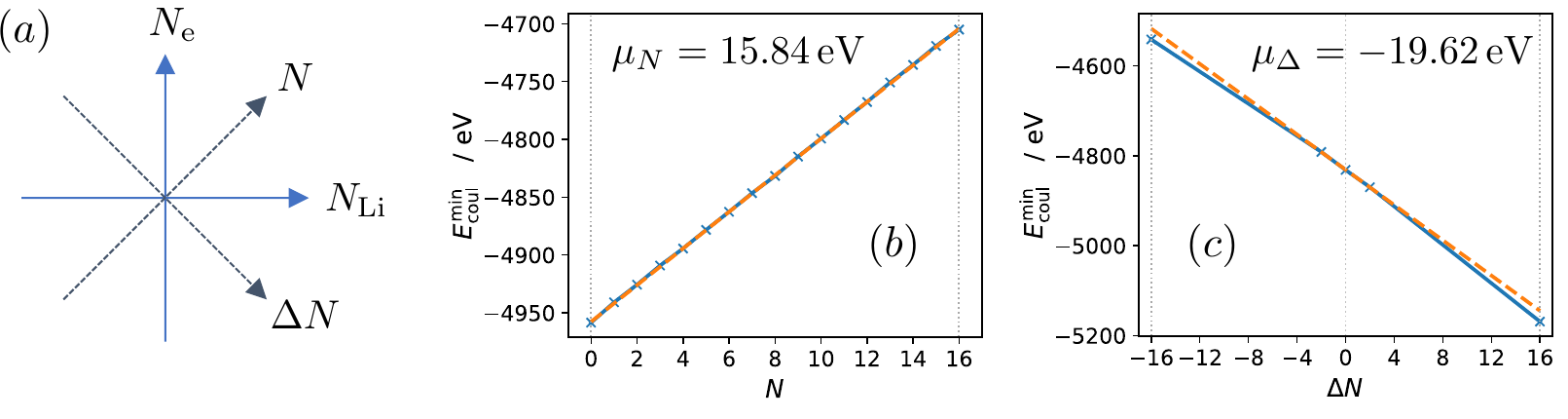}
\caption{(a) Variable transformation in particle numbers to define the number of charge-neutral \ce{Li^+}--\ce{e^-} pairs, $N = (N_{\ce{Li^+}} + N_{\ce{e^-}})/2$, and the number of net charges, $\Delta N = N_{\ce{Li^+}} - N_{\ce{e^-}}$. (b) Plot of the minimum Coulomb energy ($E_{\mathrm{coul}}^{\mathrm{min}}$) \textit{vs.} number of \ce{Li^+}--\ce{e^-} pairs ($N$), and (c) \textit{vs.} number of net charges ($\Delta N$), for the $1 \times 2 \times 2$ LFP cell, with fixed $\Delta N = 0$ in (b) and fixed $N = 8$ in (c). The chemical potentials, $\mu_{N}$ and $\mu_{\Delta}$, are determined as the slopes of the respective linear fits (orange dashed lines) around the target values for $N$ and $\Delta N$.}
\label{fig_chem_pots}
\end{figure*}

\subsection*{Electro-ionic optimization}

As explained in the Methods section, the Coulomb energy is naturally expressed as a quadratic function of the site occupation variables $x_i\in\left\{0,1\right\}$ and $y_k\in\left\{0,1\right\}$ for Li sites (\ce{Li^+}) and Fe redox centers (\ce{e^-}), respectively, \textit{cf.} Eqs.~\eqref{eq_Ecoul_Li_e} and \eqref{eq_Ecoul_Li_e_QUBO}. The task for identifying the respective minimum energy configuration thus represents a QUBO problem, which is suitable for the use of the D-Wave quantum annealing system. Since both the Li ions and the compensating redox electrons are explicitly accounted for, the QUBO model enables the simulation of LFP at different degrees of lithiation (state of charge, SOC), \textit{i.e.}, for different numbers of Li ions, $N_{\ce{Li^+}}$, and redox electrons, $N_{\ce{e^-}}$. Due to the required charge neutrality, the number of \ce{Li^+} and \ce{e^-} in the simulation cell must be matched, $N_{\ce{Li^+}} = N_{\ce{e^-}}$. In the following, the configurational ground state is explored for the $1\times2\times2$ and $1\times4\times1$ LFP cells, both comprising a total of 16 \ce{LiFePO4} units (16 Li sites and 16 Fe redox centers). Three different SOC with stoichiometries \ce{Li_{0.25}FePO4}, \ce{Li_{0.5}FePO4}, and \ce{Li_{0.75}FePO4} are targeted, corresponding to $N_{\ce{Li^+}} = N_{\ce{e^-}} = 4$, $8$, and $12$, respectively.

\subsubsection*{Particle numbers and chemical potentials}

As described in the Introduction section, a given target number of occupied sites (Hamming weight, Eq.~\eqref{eq_Hamming_weight}) is achieved by adding a soft constraint in the form of an energy penalty (Eq.~\eqref{eq_energy_penalty}) to the QUBO cost function. In the prior work~\cite{binningerOptimizationIonicConfigurations2024}, it was shown that this method is rendered ineffective if the $E$ \textit{vs.} $N$ curve has a pronounced slope ($\mu = \partial E/\partial N$), which can be mitigated by performing a Legendre transformation of the energy. The resulting grand-canonical cost function (Eq.~\eqref{eq_grand_canonical_cost_function}) becomes flat around the target particle number, which significantly facilitates the implementation of the constraint. This method is now applied to the correlated constraints in the numbers of \ce{Li^+} and \ce{e^-} for different SOC of LFP. To reflect the charge-neutrality constraint, a variable transformation is performed in the space of \ce{Li^+} and \ce{e^-} particle numbers,
\begin{align}
N & = \frac{N_{\ce{Li^+}} + N_{\ce{e^-}}}{2} \label{eq_N}\\
\Delta N & = N_{\ce{Li^+}} - N_{\ce{e^-}} \ ,
\label{eq_DeltaN}
\end{align}
defining the number of neutral \ce{Li^+}--\ce{e^-} pairs, $N$, and the number of net charges, $\Delta N$, respectively, as visualized in Fig.~\ref{fig_chem_pots}a. Changes in the SOC of LFP correspond to charge-neutral variations in $N$, while $\Delta N = 0$ remains fixed due to charge neutrality. The minimum Coulomb energy ($E_{\mathrm{coul}}^{\mathrm{min}}$) is plotted \textit{vs.} $N$ (with fixed $\Delta N = 0$) in Fig.~\ref{fig_chem_pots}b and \textit{vs.} $\Delta N$ (with fixed $N = 8$) in Fig.~\ref{fig_chem_pots}c for the $1 \times 2 \times 2$ LFP cell, where $E_{\mathrm{coul}}^{\mathrm{min}}$ was determined by full enumeration of all configurations. The chemical potentials are defined as the respective local slopes, $\mu_{N} = \partial E_{\mathrm{coul}}^{\mathrm{min}}/\partial N$ and $\mu_{\Delta} = \partial E_{\mathrm{coul}}^{\mathrm{min}}/\partial \Delta N$, at the target values for $N$ and $\Delta N$, and determined from linear fits (orange dashed lines). A perfectly linear $E$ \textit{vs.} $N$ curve is observed in Fig.~\ref{fig_chem_pots}b corresponding to a constant value of $\mu_{N} = 15.84\,\mathrm{eV}$ for any $N$. The curve in Fig.~\ref{fig_chem_pots}c has a local slope of $\mu_{\Delta} = -19.62\,\mathrm{eV}$ around $\Delta N = 0$ and a slightly negative curvature. These values correspond to chemical potentials of $\mu_{\ce{Li^+}} = \mu_{N}/2 + \mu_{\Delta} = -11.70\,\mathrm{eV}$ and $\mu_{\ce{e^-}} = \mu_{N}/2 - \mu_{\Delta} = 27.54\,\mathrm{eV}$. It should be noted that, while charged configurations with $\Delta N \neq 0$ are physically forbidden and would result in a divergence of the energy (per cell), the Coulomb model allows variations in $\Delta N$ due to an unphysical compensating background charge that is implicitly included in the Ewald energies of charged configurations. From a fundamental perspective, the positive value of $\mu_{N}$ might appear surprising, since the fully lithiated phase is generally expected to be lower in energy than the delithiated one. This seeming discrepancy is a result of the missing on-site contribution to the chemical potential of the redox electrons. In the point-charge Coulomb model, the interaction energy between a redox electron and the Fe(III) core of the same site (\textit{i.e.} same position) would amount to minus infinity and is therefore excluded in the summation of Coulomb terms (\textit{cf.} Eq.~\eqref{eq_Ecoul_Li_e} and discussion thereafter). The next-order leading term in the electronic chemical potential is positive, producing an overall positive value of $\mu_{N}$. This situation is different, \textit{e.g.}, in DFT calculations where the electronic charge is distributed in orbitals with a finite negative on-site energy. For the purpose of the present study, however, such differences are irrelevant, because the on-site contribution does not affect the relative energies of different configurations with the same $N$. Any differences due to sign and value of the chemical potential get cancelled after performing the respective Legendre transformation to the grand-canonical cost function, \textit{cf.} Eq.~\eqref{eq_total_QUBO}.

\begin{figure*}[t]
\centering
\includegraphics[width=0.95\textwidth]{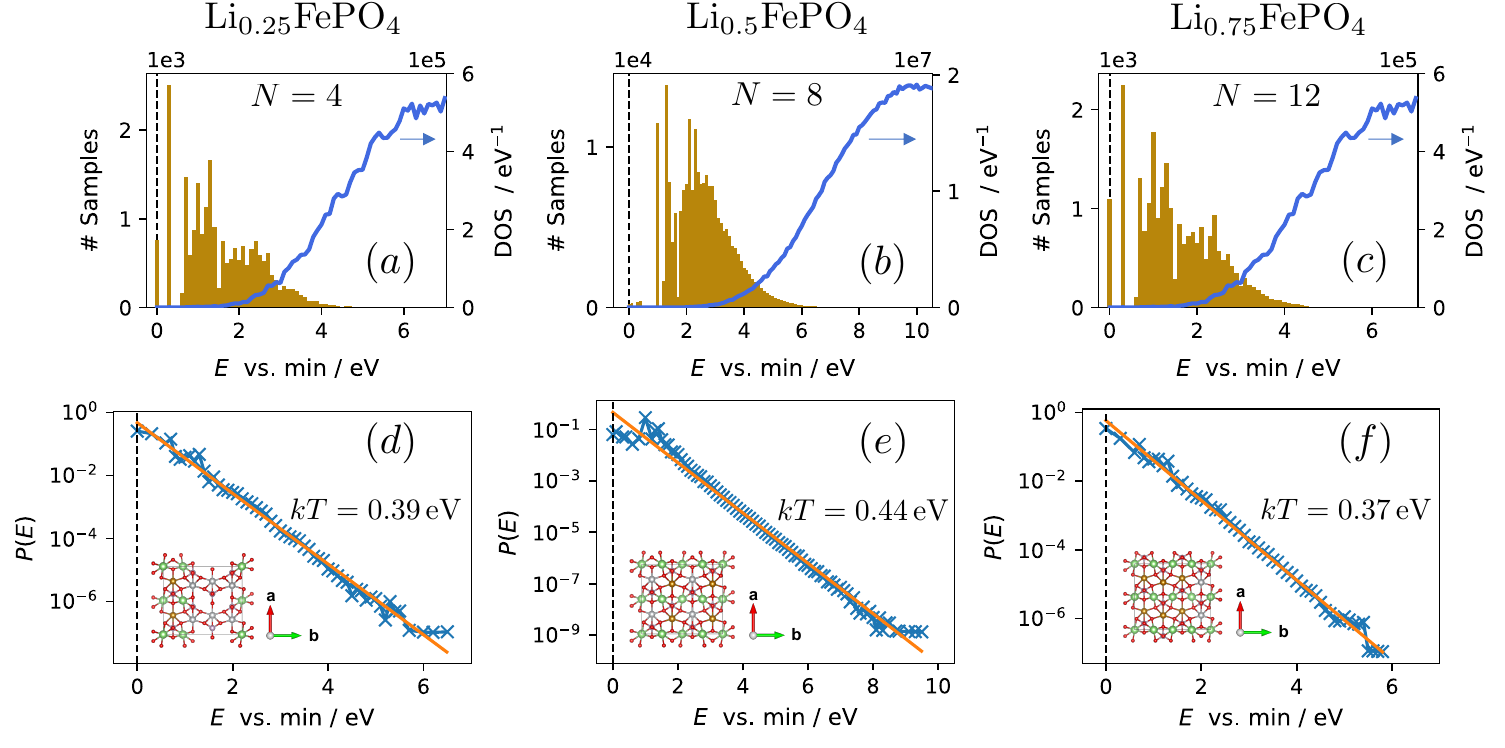}
\caption{(a)--(c) Histogram of the energies of QA-sampled configurations (golden bars) and density of states (DOS) of the Coulomb model (blue curve) as a function of the energy, relative to the minimum energy in the ground state, for the $1 \times 2 \times 2$ LFP cell. (d)--(f) Normalized sampling probability $P(E)$, plotted on a logarithmic scale, and Boltzmann fit with given effective sampling temperatures $kT$. The respective ground-state configurations are shown as insets. SOC of $N=4$ in (a) and (d), $N=8$ in (b) and (e), and $N=12$ in (c) and (f) for a total of 16 \ce{Li^+} and \ce{e^-} sites.}
\label{fig_QA_LFP_1x2x2}
\end{figure*}

\begin{table}[tbp]
  \centering
  \caption{QA parameters and output statistics on D-Wave Advantage\texttrademark{} 5.4 System. In all cases, the values of the chemical potentials were fixed at $\mu_{N} = 15.84\,\mathrm{eV}$ and $\mu_{\Delta} = -19.62\,\mathrm{eV}$, and the annealing time was $t_{\mathrm{ann}} = 100\,\mathrm{\mu s}$. c-str.: chain strength; $E_{\mathrm{min}}^{\mathrm{QA}}$: minimum Coulomb energy returned by QA; $E_{\mathrm{min}}^{\mathrm{glob}}$: global minimum of the Coulomb energy for the given cell dimensions and stoichiometries determined by full enumeration. $\%_{E_{\mathrm{min}}^{\mathrm{glob}}}^{\mathrm{QA}}$: QA output fidelity (fraction of optimal solutions).}
    \begin{tabular}{c|c|c|c|c|c|c|c|c|c}
    Cell  & Composition & $N^{\mathrm{target}}$ & $\lambda_{N}\,/\,\mathrm{eV}$ & $\lambda_{\Delta}\,/\,\mathrm{eV}$ & c-str. & $E_{\mathrm{min}}^{\mathrm{QA}}\,/\,\mathrm{eV}$ & $E_{\mathrm{min}}^{\mathrm{glob}}\,/\,\mathrm{eV}$ & No. runs & $\%_{E_{\mathrm{min}}^{\mathrm{glob}}}^{\mathrm{QA}}$ \\
    \hline
    $1\times 2\times 2$ & \ce{Li_{0.25}FePO4} & 4     & 0.4   & 0.4   & 2     & $-$4894.394 & $-$4894.394 & 100\,000 & 0.76\% \\
    $1\times 2\times 2$ & \ce{Li_{0.5}FePO4} & 8     & 0.4   & 0.4   & 2     & $-$4831.699 & $-$4831.699 & 800\,000 & 0.03\% \\
    $1\times 2\times 2$ & \ce{Li_{0.75}FePO4} & 12    & 0.4   & 0.4   & 2     & $-$4767.677 & $-$4767.677 & 100\,000 & 1.09\% \\
    \hline
    $1\times 2\times 2$ & \ce{Li_{0.25}FePO4} & 4     & 0     & 0.4   & 2     & $-$4894.394 & $-$4894.394 & 500\,000 & 0.09\% \\
    $1\times 2\times 2$ & \ce{Li_{0.5}FePO4} & 8     & 0     & 0.4   & 2     & $-$4831.699 & $-$4831.699 & 500\,000 & 0.002\% \\
    $1\times 2\times 2$ & \ce{Li_{0.75}FePO4} & 12    & 0     & 0.4   & 2     & $-$4767.677 & $-$4767.677 & 500\,000 & 0.23\% \\
    \hline
    $1\times 4\times 1$ & \ce{Li_{0.25}FePO4} & 4     & 1     & 1     & 8     & $-$4894.304 & $-$4894.304 & 100\,000 & 0.02\% \\
    $1\times 4\times 1$ & \ce{Li_{0.5}FePO4} & 8     & 1     & 1     & 8     & $-$4830.373 & $-$4831.280 & 100\,000 & 0\% \\
    $1\times 4\times 1$ & \ce{Li_{0.75}FePO4} & 12    & 1     & 1     & 8     & $-$4767.585 & $-$4767.585 & 100\,000 & 0.02\% \\
    \hline
    $1\times 4\times 1$ & \ce{Li_{0.25}FePO4} & 4     & 0     & 1     & 8     & $-$4894.304 & $-$4894.304 & 500\,000 & 0.0004\% \\
    $1\times 4\times 1$ & \ce{Li_{0.5}FePO4} & 8     & 0     & 1     & 8     & $-$4830.373 & $-$4831.280 & 500\,000 & 0\% \\
    $1\times 4\times 1$ & \ce{Li_{0.75}FePO4} & 12    & 0     & 1     & 8     & $-$4767.585 & $-$4767.585 & 500\,000 & 0.004\% \\
    \end{tabular}%
  \label{tab_QA_LFP}
\end{table}%

\subsubsection*{Quantum annealing}

The total QUBO cost function used for the quantum annealing procedure reads
\begin{align}
E_{\mathrm{cost}}(\{x_i\},\{y_k\})\, =\, E_{\mathrm{coul}} \, -\mu_{\Delta}\Delta N \, -\mu_{N} N \, +\lambda_{\Delta}\left(\Delta N - \Delta N^{\mathrm{target}} \right)^2 \, +\lambda_{N}\left( N - N^{\mathrm{target}} \right)^2 \ ,
\label{eq_total_QUBO}
\end{align}
where $E_{\mathrm{coul}}$ is the Coulomb energy as a function of $x_i$ and $y_k$ (Eq.~\eqref{eq_Ecoul_Li_e_QUBO}), and $N$ and $\Delta N$ are given by Eqs.~\eqref{eq_N} and \eqref{eq_DeltaN}, respectively, with $N_{\ce{Li^+}} = \sum_i x_i$ and $N_{\ce{e^-}} = \sum_k y_k$. The Legendre-transformed energy, $E_{\mathrm{coul}} \ -\mu_{\Delta}\Delta N \ -\mu_{N} N$, has zero slopes in $N$ and $\Delta N$, for which reason the quadratic energy penalty terms with small values of the parameters $\lambda_{\Delta}$ and $\lambda_{N}$ are sufficient to produce a global minimum of the total cost function~\eqref{eq_total_QUBO} at the target values for $\Delta N$ and $N$. Due to the charge-neutrality requirement, the former was fixed at $\Delta N^{\mathrm{target}} = 0$, while $N^{\mathrm{target}}$ was varied according to the targeted SOC. The definition of the chemical potentials, $\mu_{N}$ and $\mu_{\Delta}$, was described in the previous section. In practice, however, the use of full enumeration or classical heuristics for the determination of the required $E$ \textit{vs.} $N$ or $\Delta N$ curves would not be desirable as part of a quantum optimization method. A simplified method is thus suggested, which consists in approximating the chemical potentials in terms of the finite differences between the end points of the $E^{\mathrm{min}}$ \textit{vs.} $N$ / $\Delta N$ curves. Since the end-point configurations with $N=0,\,16$ and $\Delta N = -16,\,16$ in Fig.~\ref{fig_chem_pots}b and c, respectively, are trivial, meaning that they comprise either fully occupied or entirely vacant sublattices for \ce{Li^+} and \ce{e^-}, the respective energies are readily calculated. This method yields values of $\mu_{N} = 15.84\,\mathrm{eV}$ and $\mu_{\Delta} = -19.62\,\mathrm{eV}$, identical to the values determined by local fits, \textit{cf.} Fig.~\ref{fig_chem_pots}b,c.

The results of the configurational sampling by quantum annealing on the D-Wave Advantage\texttrademark{} 5.4 System are presented in Fig.~\ref{fig_QA_LFP_1x2x2} for the case of the $1 \times 2 \times 2$ LFP cell at $25\%$, $50\%$, and $75\%$ lithiation degree, corresponding to $N^{\mathrm{target}} = 4$ (Fig.~\ref{fig_QA_LFP_1x2x2}a,d), $N^{\mathrm{target}} = 8$ (Fig.~\ref{fig_QA_LFP_1x2x2}b,e), and $N^{\mathrm{target}} = 12$ (Fig.~\ref{fig_QA_LFP_1x2x2}c,f). The respective parameters and output statistics of the quantum-annealing procedure are summarized in Table~\ref{tab_QA_LFP} (first block). In Fig.~\ref{fig_QA_LFP_1x2x2}a--c, the QA histograms of sampled energies (golden bars) are plotted together with the total density of states (DOS) of the Coulomb model determined by full enumeration (blue curve). While the sampled energies are clustered in the lower-energy range of the spectrum, the highest sampling frequencies are not observed close to the optimal solutions, \textit{viz.} the minimum energy. The respective fractions of optimal solutions returned by QA (\textit{a.k.a.} QA fidelity) are presented in Table~\ref{tab_QA_LFP} ($\%_{E_{\mathrm{min}}^{\mathrm{glob}}}^{\mathrm{QA}}$). While for $N=4$ and $N=8$, the ground state was returned in about $1\%$ of annealing runs, the fraction decreased to about $0.03\%$ for $N=8$ due to the much larger combinatorial subspace ($165\,636\,900$ possible configurations for $N=8$ compared to $3\,312\,400$ for $N=4$ and $8$). However, as discussed previously~\cite{binningerOptimizationIonicConfigurations2024}, the apparent sampling frequency (per energy) is given by the intrinsic sampling probability, $P(E)$, weighted with the density of states ($g_{\mathrm{DOS}}(E)$) of the QUBO model~\cite{bruggerOutputStatisticsQuantum2022},
\begin{align}
N_{\mathrm{Samples}}(E) \,\propto\, P(E)\,g_{\mathrm{DOS}}(E) \ .
\end{align}
The probability $P(E)$ was thus estimated by normalizing the QA output histogram with the respective DOS, and the result is presented in Fig.~\ref{fig_QA_LFP_1x2x2}d--f on a logarithmic scale. As for the previously studied case of LCO~\cite{binningerOptimizationIonicConfigurations2024}, an exponentially decreasing intrinsic sampling probability is observed, which is characterized by an effective sampling temperature, $P(E) \propto \exp(-E/kT)$. In all cases, a fitted value of about $kT = 0.4\,\mathrm{eV}$ was found. Importantly, the Boltzmann-type statistics imply a maximum likelihood for sampling an optimal solution. For the semi-lithiated stoichiometry ($N=8$), however, the lowest-energy range had a slightly decreased probability in comparison to the Boltzmann fit (Fig.~\ref{fig_QA_LFP_1x2x2}e). The same behavior is observed when analyzing any subset of the respective $800\,000$ annealing runs, meaning that the undersampling was a systematic effect and not a result of statistical variations due to a limited number of annealing runs.

\begin{figure*}[t]
\centering
\includegraphics[width=0.95\textwidth]{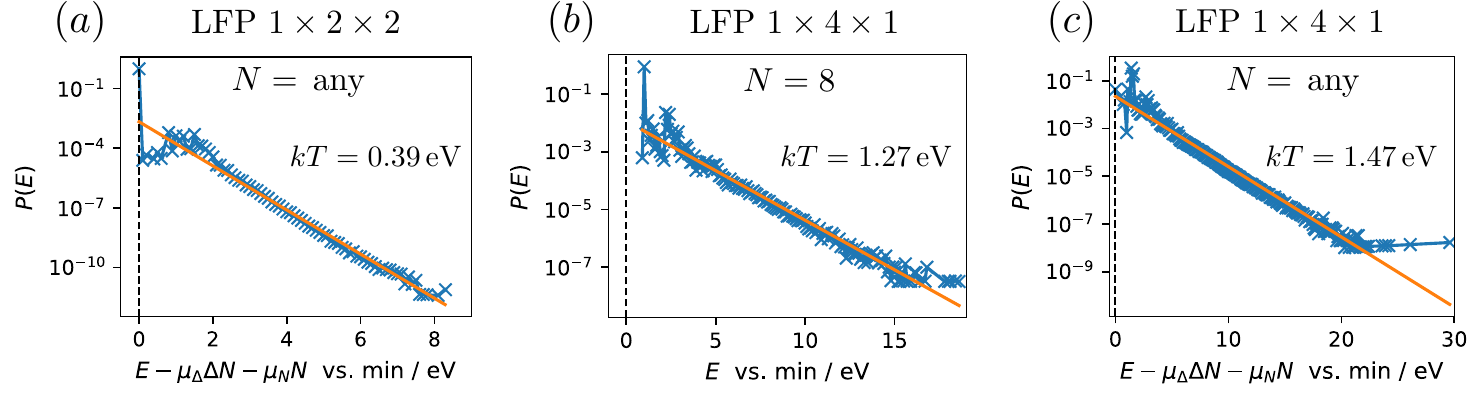}
\caption{Normalized sampling probability $P(E)$ (plotted on a logarithmic scale) and Boltzmann fit with given effective sampling temperature $kT$, for (a) the $1 \times 2 \times 2$ LFP cell with $N$-unconstrained cost function ($\lambda_{N} = 0$), (b) the $1 \times 4 \times 1$ LFP cell with $N = 8$ constrained cost function ($\lambda_{N} = 1\,\mathrm{eV}$), and (c) the $1 \times 4 \times 1$ LFP cell with $N$-unconstrained cost function ($\lambda_{N} = 0$).}
\label{fig_QA_LFP_1x2x2_1x4x1}
\end{figure*}

In the QA results discussed so far, a certain target stoichiometry ($N^{\mathrm{target}}$) was enforced by a positive value of the constraint parameter $\lambda_{N} = 0.4\,\mathrm{eV}$, \textit{cf.} first block in Table~\ref{tab_QA_LFP}, the value of which was optimized to maximize the fraction of sampled configurations that fulfilled the constraint, while maintaining a minimum possible value of the chain strength parameter (c-str. in Table~\ref{tab_QA_LFP}) required to prevent breakages of the physical qubit chains that represent each logical variable. On the other hand, due to the nearly perfect linearity of the energy \textit{vs.} particle number curve shown in Fig.~\ref{fig_chem_pots}b, the Legendre-transformed energy ($E_{\mathrm{coul}} -\mu_{N} N$) becomes flat across the entire range of $N$. Setting $\lambda_{N} = 0$ in the cost function~\eqref{eq_total_QUBO}, thus implies essentially identical values of the cost minimum at any $N$. It is therefore expected that QA sampling with $\lambda_{N} = 0$ will sample the ground-state configurations for different SOC at similar frequencies. Indeed, as shown by the results presented in the second block of Table~\ref{tab_QA_LFP}, $500\,000$ QA runs with $\lambda_{N} = 0$ returned the minimum energy configurations for the different stoichiometries, albeit with different frequencies. The intrinsic sampling probability for the $N$-unconstrained cost function, shown in Fig.~\ref{fig_QA_LFP_1x2x2_1x4x1}a, is characterized by the same Boltzmann-type statistics as for the $N$-constrained method (Fig.~\ref{fig_QA_LFP_1x2x2}d--f), with a similar effective sampling temperature of $kT = 0.39\,\mathrm{eV}$. Couzini\'e et al.~\cite{couzinieAnnealingPredictionGrand2024} recently presented a similar unconstrained `grand-canonical' optimization method for the prediction of crystal structures by quantum annealing, however, without involving a Legendre transformation in particle numbers. It is expected that the present method will be readily applicable to such related problems.

As a second system, the $1 \times 4 \times 1$ LFP supercell was studied by the same approach to investigate the influence of the cell geometry on the QA statistics. The $1 \times 4 \times 1$ cell has the same total number of 16 sites for \ce{Li^+} and \ce{e^-} at a larger asymmetry of the cell parameters ($10.332 \times 24.040 \times 4.692 \ \text{\AA}^3$) as compared to the more regular $1 \times 2 \times 2$ cell ($10.332 \times 12.020 \times 9.384 \ \text{\AA}^3$). Interestingly, the $1 \times 4 \times 1$ geometry yielded identical values for the chemical potentials as for the $1 \times 2 \times 2$ cell, $\mu_{N} = 15.84\,\mathrm{eV}$ and $\mu_{\Delta} = -19.62\,\mathrm{eV}$. The respective QA parameters and output statistics are complied in the third and fourth blocks of Table~\ref{tab_QA_LFP}. Despite the identical size of the combinatorial space, the QA method was significantly less effective for identifying the ground-state configurations in the asymmetric cell, as quantified by very small fidelities ($\%_{E_{\mathrm{min}}^{\mathrm{glob}}}^{\mathrm{QA}}$). In particular, the global optimum for the $N = 8$ problem was not returned once by the QA procedure. This is reflected in the intrinsic sampling probabilities presented in Fig.~\ref{fig_QA_LFP_1x2x2_1x4x1}b,c. Both for the $N = 8$ constrained optimization with $\lambda_{N} = 1\,\mathrm{eV}$ (Fig.~\ref{fig_QA_LFP_1x2x2_1x4x1}b) and the $N$-unconstrained sampling with $\lambda_{N} = 0$ (Fig.~\ref{fig_QA_LFP_1x2x2_1x4x1}c), Boltzmann-type statistics are observed, but with large effective sampling temperatures of about $kT = 1.3$--$1.5\,\mathrm{eV}$. The dependence of the QA effectiveness on the supercell geometry (for same combinatorial problem size) can be related to the variance of the respective QUBO coefficients. For the $1 \times 2 \times 2$ system, the distribution of the off-diagonal Coulomb coefficients is characterized by a standard deviation of $0.69\,\mathrm{eV}$ with maximum and minimum values of $1.09\,\mathrm{eV}$ and $-1.10\,\mathrm{eV}$, respectively, whereas the coefficients for the $1 \times 4 \times 1$ system have a much wider distribution with a standard deviation of $2.88\,\mathrm{eV}$ and maximum/minimum values of $3.78\,\mathrm{eV}$ and $-4.65\,\mathrm{eV}$. The larger variance of the coefficients for the latter system is due to the asymmetrically elongated shape of the cell which accommodates larger variations in the spatial separation between the charged species. It is noted that the factor of about $4$ between the spreads of QUBO coefficients is similar to the relative magnitude of the effective QA sampling temperatures for the two different cell geometries.

\section*{Discussion}

The Boltzmann-type statistics is a general feature of QA on practical D-Wave devices as reported in several studies~\cite{aminSearchingQuantumSpeedup2015, benedettiEstimationEffectiveTemperatures2016, bruggerOutputStatisticsQuantum2022}. While generically, QA is devised to search for the ground-state solution of an optimization problem, the thermal output statistics make the use of QA also interesting for problems where the target consists in thermal averages over the configuration space~\cite{benedettiQuantumAssistedLearningHardwareEmbedded2017, sandtEfficientLowTemperature2023}. However, Benedetti \textit{et al.}~\cite{benedettiEstimationEffectiveTemperatures2016} pointed out that the effective sampling temperature is dependent on the problem at hand, also shown in the present results (Figs.~\ref{fig_QA_LFP_1x2x2} and \ref{fig_QA_LFP_1x2x2_1x4x1}), and not defined by the physical temperature of the QA hardware, which is in the range of milli-Kelvin. Decoherence and noise during the QA process is considered to contribute to the output statistics. Brugger \textit{et al.}~\cite{bruggerOutputStatisticsQuantum2022} demonstrated that the Boltzmann-like output could be explained by a finite precision in the tuning of the target Hamiltonian parameters. While the effective sampling temperature of the QA process cannot be controlled, the Boltzmann statistics provide access to thermodynamically relevant low-energy configurations, which was used by Sandt and Spatschek~\cite{sandtEfficientLowTemperature2023} for post-estimation of thermodynamic quantities at selected target temperatures.

\begin{figure*}[t]
\centering
\includegraphics[width=0.95\textwidth]{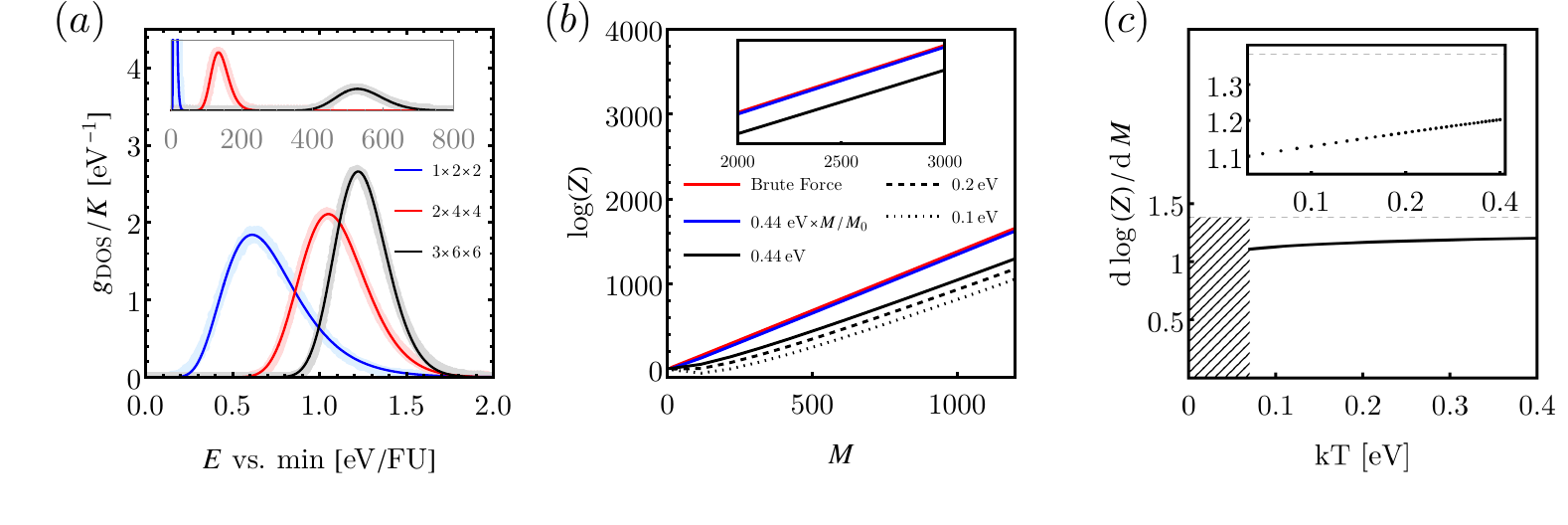}
\caption{(a) Joint density of states ($g_{\mathrm{DOS}}$, normalized by the total number of states $K$) of the electro-ionic Coulomb model for the semi-lithiated $1\times2\times2$, $2\times4\times4$, and $3\times6\times6$ LFP cells, determined by random sampling (shaded lines), plotted versus the energy per formula unit (FU) with the ground-state energy as zero. Log-normal fits are shown as solid lines. Inset: Same plots on an absolute energy scale (per supercell). (b) Logarithm (natural) of the partition function (Eq.~\eqref{eq_partition_function}) as a function of the supercell size $M$ (number of LFP formula units). Inset: Same plot for larger values of $M$. (c) Derivative (slope) of the exponent of $Z$ with respect to the system size ($M$), plotted as a function of the effective sampling temperature $kT$. The calculation failed for numerics reasons in the range below $0.1\,\mathrm{eV}$ (shaded area). The reference value for the brute-force method (horizontal dashed line), $\mathrm{d}\log(K)/\mathrm{d}M = \log(4) \approx 1.39$, follows from the exponential behavior $K\propto 4^M$ obtained by using Stirling's approximation. Inset: Enlarged view of the plot.}
\label{fig_DOS}
\end{figure*}

The target in the present work was the identification of an optimum solution, \textit{i.e.}, a lowest-energy configuration of the electro-ionic Hamiltonian. An ideal QA process would return an optimum with deterministic certainty for any single run, and thus provide an exponential speed-up compared to classical deterministic methods, which effectively require extensive exploration (full enumeration) of the complete configuration space. The thermal output distributions observed in practice, however, make QA a quantum heuristic method, raising questions on practical limitations in optimization performance and scaling behavior. The quantum computation time needed for the probabilistic identification of an optimum solution, $t_{\mathrm{qc}} = t_{\mathrm{ann}}/P_0$, corresponds to the annealing time of a single QA run multiplied by the inverse of the success probability ($P_0$) to return a ground-state configuration~\cite{aminSearchingQuantumSpeedup2015}, which corresponds to the number of QA runs required to achieve a unity expectation value for a successful outcome. In the following, an empirical estimate is made for the scaling of $t_{\mathrm{qc}}$ as a function of the size of the problem instance, namely, the size of the LFP model cell. To this end, the intrinsic sampling probability is assumed to follow a Boltzmann distribution, $P(E) = \exp(-E/kT)/Z$, with a certain effective sampling temperature $kT$, as observed for the actual QA results across a major part of the output spectrum, \textit{cf.} Figs.~\ref{fig_QA_LFP_1x2x2} and \ref{fig_QA_LFP_1x2x2_1x4x1}. Since the probability must integrate to unity, the normalization constant $Z$ is given by the effective partition function,
\begin{align}
Z = \int_0^{\infty} e^{-E/kT}\,g_{\mathrm{DOS}}(E)\,\mathrm{d}E \ ,
\label{eq_partition_function}
\end{align}
with the density of states $g_{\mathrm{DOS}}(E)$, and the ground-state energy taken as zero. With a ground-state degeneracy (multiplicity) of $g_0$, the success probability is given by $P_0 = g_0/Z$ and the total computation time is thus proportional to the partition function,
\begin{align}
t_{\mathrm{qc}} = t_{\mathrm{ann}}\,\frac{Z}{g_0} \ .
\label{eq_qc_time}
\end{align}
Fig.~\ref{fig_DOS}a shows the DOS for the semi-lithiated $1\times2\times2$, $2\times4\times4$, and $3\times6\times6$ LFP cells, as determined by random sampling (shaded lines). In all cases, the DOS were well fitted by log-normal distributions (solid lines),
\begin{align}
g_{\mathrm{DOS}}^{\mathrm{fit}}(E) = \frac{K}{\sqrt{2\pi}\,\sigma\,E}\,\exp\left(-\frac{\log^2(E/\epsilon)}{2\sigma^2}\right) \ ,
\label{eq_lognorm_DOS}
\end{align}
where $K = \binom{M}{M/2}^2$ is the total number of configurations for cells with $M$ formula units of \ce{LiFePO4}, \textit{i.e.}, $M$ sites for both \ce{Li^+} and \ce{e^-}, each of which being half occupied. The fitted parameters were found to depend on the cell size, with $\sigma = 0.33$, $0.18$, and $0.12$, and $\epsilon = 11.0$, $139$, and $536\,\mathrm{eV}$ for the $1\times2\times2$ (with $M = 16$), $2\times4\times4$ (with $M = 128$), and $3\times6\times6$ cells (with $M = 432$), respectively. In Eq.~\eqref{eq_lognorm_DOS}, $E$ refers to the total energy for each cell, while in Fig.~\ref{fig_DOS}a $E$ is given per formula unit. 

This empirical DOS model can be used to estimate the scaling behavior of the quantum computation time~\eqref{eq_qc_time}, \textit{viz.} the effective partition function~\eqref{eq_partition_function}, as a function of $M$. It is assumed that the respective DOS can be described by the log-normal model~\eqref{eq_lognorm_DOS} with parameter functions, $\sigma (M) = 0.13 + 3.29/M$ and $\epsilon (M) = (-9.22 + 1.26\,M)\,\mathrm{eV}$, whereby the roughly linear dependency of the peak-maximum energy on the system size can be explained by the extensive nature of the total energy. Fig.~\ref{fig_DOS}b shows the respective scaling of the logarithm of $Z$ (Eq.~\eqref{eq_partition_function}) with $M$ for different values of the effective sampling temperature. Besides the fixed values of $kT=0.1$, $0.2$, and $0.44\,\mathrm{eV}$ (the latter value taken as a representative for the fitted values in Fig.~\ref{fig_QA_LFP_1x2x2}), a sampling model with an effective temperature $kT=0.44\,\mathrm{eV}\times M/M_0$ linearly increasing with system size was considered (with $M_0 = 16$ corresponding to the $1 \times 2 \times 2$ cell). The latter model (blue curve in Fig.~\ref{fig_DOS}b) showed essentially identical scaling behaviour as brute-force sampling, \textit{viz.} full enumeration, with $Z = K$ equal to the total number of configurations (red curve in Fig.~\ref{fig_DOS}b). In contrast, the models with a fixed QA sampling temperature yielded significantly decreased values of $\log(Z)$. The slope of the scaling plots ($\mathrm{d}\log(Z)/\mathrm{d}M$), determined in the range $M=500$--$1000$, is shown in Fig.~\ref{fig_DOS}c as a function of the (fixed) effective sampling temperature. Slope values of about $1.1$--$1.2$ were obtained, which decreased with decreasing sampling temperature, and which are lower than the reference value for the brute-force method, $\mathrm{d}\log(K)/\mathrm{d}M = \log(4) \approx 1.39$ (horizontal dashed line in Fig.~\ref{fig_DOS}c). While such lower slope values would indicate an exponential speed-up provided by the sampling method, it should be noted that the observed slopes become parallel to the brute-force reference at very large $M$ (inset in Fig.~\ref{fig_DOS}b), eventually resulting in the same exponential scaling of the computation time with system size. These estimations indicate that, in a strict sense, an exponential speed-up might be challenging to achieve with thermal sampling statistics. 

It is important to note that such considerations only refer to the scaling of the exponent of $Z$ as a function of system size. The respective pre-exponential factors are much smaller than for the brute-force reference, as shown by the offsets between the curves in Fig.~\ref{fig_DOS}b. This would translate into significant speed-up in total computation time, which could lead to practical advantages of the QA method for the solution of coulombic optimization problems in the future.

\section*{Conclusion}

The joint optimization of the ground-state distributions of lithium ions and redox electrons in a model cell of lithium iron phosphate was studied based on an electro-ionic point-charge Coulomb model. Using classical heuristic optimization, the model was shown to correctly predict the LFP charging characteristics, namely the bulk \ce{LiFePO4}/\ce{FePO4} phase separation with a preferred phase interface in the $ac$ plane, and a solid-solution behavior for particles below about $40\,\mathrm{nm}$. A method was presented for performing the ground-state search by quantum annealing. To this end, the cost function was Legendre-transformed using the chemical potentials of charge-neutral (physically allowed) and charged (artificial) variations in lithium ion and electron numbers. This method enabled the successful identification of the lowest-energy configuration in an LFP model cell at different state of charge on D-Wave quantum-annealing hardware. After normalization with the electro-ionic density of states, the QA sampling output was found to follow approximately Boltzmann-type statistics, the consequences of which for the scaling of the quantum computation time with system size were estimated. This work takes a step towards the practical use of quantum computers for joint quantum simulation and combinatorial optimization of electrons and ions in energy materials.

\section*{Methods}

\subsection*{Structural model}

LFP simulation cells were constructed from the orthorhombic unit cell of triphylite \ce{LiFePO4} with lattice constants $a=10.332\,\text{\AA}$, $b=6.010\,\text{\AA}$, and $c=4.692\,\text{\AA}$ (comprising 4 formula units of \ce{LiFePO4}), as provided in the crystallographic information file (CIF) No. 2100916 of the Crystallography Open Database (COD) according to Streltsov \textit{et al.}~\cite{streltsovMultipoleAnalysisElectron1993}. Two different supercells, $1\times2\times2$ and $1\times4\times1$, were considered for the configurational optimization, comprising a total of 16 formula units of \ce{LiFePO4} with 16 sites of both Li and Fe, \textit{cf.} Fig.~\ref{fig_LFP_model}a. The cell geometry was frozen for all degrees of lithiation, \textit{i.e.}, the contraction of the unit cell volume by about $7\%$ for the isostructural \ce{FePO4} heterosite end-member~\cite{anderssonLithiumExtractionInsertion2000} was neglected. Structural drawings were generated using the VESTA software~\cite{mommaVESTA3Threedimensional2011}.

\subsection*{Energy model}

As in the previous work~\cite{binningerOptimizationIonicConfigurations2024}, a point-charge Coulomb interaction model was employed to describe the energy of different \ce{Li^+} and \ce{e^-} (\ce{Fe(II)}/\ce{Fe(III)}) configurations in the LFP simulation cell,
\begin{align}
E_{\mathrm{coul}} = \frac{e^2}{4\pi\epsilon_0}\,\sum_{\alpha<\beta} \frac{Z_{\alpha}\,Z_{\beta}}{|\vec{r}_{\alpha}-\vec{r}_{\beta}|} \ ,
\end{align}
where the indices $\alpha$ and $\beta$ run over all ions of a given configuration. Formal valencies of $Z_{\ce{Li}} = +1$, $Z_{\ce{P}} = +5$, $Z_{\ce{O}} = -2$ were used, and either $Z_{\ce{Fe(II)}} = +2$ or $Z_{\ce{Fe(III)}} = +3$ for iron cations with a mixed valency. Site occupation variables $x_i\in\left\{0,1\right\}$ are defined to indicate whether a given Li site is occupied ($x_i = 1$) or vacant ($x_i = 0$). Likewise, binary variables $y_k\in\left\{0,1\right\}$ are introduced to describe whether a given Fe site corresponds to an \ce{Fe(II)} ($y_k = 1$) or \ce{Fe(III)} ($y_k = 0$) species. With this, the Coulomb energy of any configuration can be expressed as a function of the variables $x_i$ and $y_k$,
\begin{align}
& E_{\mathrm{coul}}(\{x_i\},\{y_k\}) = \frac{e^2}{4\pi\epsilon_0}\,\left\{\sum_{p<q\,\in\,\mathrm{fix}} \frac{Z_{p}\,Z_{q}}{|\vec{r}_{p}-\vec{r}_{q}|} + \sum_{i\,\in\,\mathcal{S}_{\mathrm{Li}}}x_i\left[\sum_{p\,\in\,\mathrm{fix}} \frac{Z_p}{|\vec{r}_i-\vec{r}_p|}\right] + \right. \nonumber \\[0.2cm] 
& \quad \left. + \sum_{k\,\in\,\mathcal{S}_{\mathrm{Fe}}}y_k\left[\sum_{p\,\in\,\mathrm{fix}} \frac{2\,Z_p}{|\vec{r}_k-\vec{r}_p|}\right] + \sum_{k\,\in\,\mathcal{S}_{\mathrm{Fe}}}(1-y_k)\left[\sum_{p\,\in\,\mathrm{fix}} \frac{3\,Z_p}{|\vec{r}_k-\vec{r}_p|}\right] + \sum_{i<j\,\in\,\mathcal{S}_{\mathrm{Li}}}x_i\,x_j\,\frac{1}{|\vec{r}_i-\vec{r}_j|}\, + \right. \nonumber \\[0.2cm] 
& \quad \left. + \sum_{\substack{i\,\in\,\mathcal{S}_{\mathrm{Li}}\\ k\,\in\,\mathcal{S}_{\mathrm{Fe}}}} x_i\,y_k\,\frac{2}{|\vec{r}_i-\vec{r}_k|}\, + \sum_{\substack{i\,\in\,\mathcal{S}_{\mathrm{Li}}\\ k\,\in\,\mathcal{S}_{\mathrm{Fe}}}} x_i\,(1-y_k)\,\frac{3}{|\vec{r}_i-\vec{r}_k|}\, + \sum_{k<l\,\in\,\mathcal{S}_{\mathrm{Fe}}}y_k\,y_l\,\frac{4}{|\vec{r}_k-\vec{r}_l|}\, + \right. \nonumber \\[0.2cm] 
& \quad \left. + \sum_{\substack{k,l\,\in\,\mathcal{S}_{\mathrm{Fe}}\\k\neq l}}y_k\,(1-y_l)\,\frac{6}{|\vec{r}_k-\vec{r}_l|}\, + \sum_{k<l\,\in\,\mathcal{S}_{\mathrm{Fe}}}(1-y_k)\,(1-y_l)\,\frac{9}{|\vec{r}_k-\vec{r}_l|} \right\} \ ,
\label{eq_Ecoul_Li_Fe2_Fe3}
\end{align}
where the first term at the right-hand side captures all interactions among the fixed species, \textit{i.e.}, phosphor and oxygen, and the second, third, and fourth terms represent interactions between fixed species and lithium, ferrous (\ce{Fe(II)}), and ferric (\ce{Fe(III)}) ions, respectively. The fifth, sixth, and seventh terms describe the interactions between \ce{Li^+}--\ce{Li^+}, \ce{Li^+}--\ce{Fe(II)}, and \ce{Li^+}--\ce{Fe(III)} pairs, and the eighth, ninth, and tenth terms represent the interactions between \ce{Fe(II)}--\ce{Fe(II)}, \ce{Fe(II)}--\ce{Fe(III)}, and \ce{Fe(III)}--\ce{Fe(III)} pairs, respectively. Regrouping and combining terms of the same order in the variables $x_i$ and $y_k$ yields
\begin{align}
& E_{\mathrm{coul}}(\{x_i\},\{y_k\}) = \frac{e^2}{4\pi\epsilon_0}\,\left\{ \sum_{p<q\,\in\,\mathrm{fix}} \frac{Z_{p}\,Z_{q}}{|\vec{r}_{p}-\vec{r}_{q}|}\, + \sum_{\substack{k\,\in\,\mathcal{S}_{\mathrm{Fe}}\\ p\,\in\,\mathrm{fix}}} \frac{3\,Z_p}{|\vec{r}_k-\vec{r}_p|}\, + \sum_{k<l\,\in\,\mathcal{S}_{\mathrm{Fe}}}\frac{9}{|\vec{r}_k-\vec{r}_l|}\, + \right.\nonumber \\[0.2cm] 
& \left. + \sum_{i\,\in\,\mathcal{S}_{\mathrm{Li}}}x_i\left[\sum_{p\,\in\,\mathrm{fix}} \frac{Z_p}{|\vec{r}_i-\vec{r}_p|} + \sum_{k\,\in\,\mathcal{S}_{\mathrm{Fe}}}\frac{3}{|\vec{r}_i-\vec{r}_k|}\right]\ +\ \sum_{k\,\in\,\mathcal{S}_{\mathrm{Fe}}}y_k\left[\sum_{p\,\in\,\mathrm{fix}} \frac{-Z_p}{|\vec{r}_k-\vec{r}_p|} + \sum_{l\neq k\,\in\,\mathcal{S}_{\mathrm{Fe}}}\frac{-3}{|\vec{r}_k-\vec{r}_l|}\right] + \right.\nonumber \\[0.2cm] 
& \left. + \sum_{i<j\,\in\,\mathcal{S}_{\mathrm{Li}}}x_i\,x_j\,\frac{1}{|\vec{r}_i-\vec{r}_j|}\, + \sum_{\substack{i\,\in\,\mathcal{S}_{\mathrm{Li}}\\ k\,\in\,\mathcal{S}_{\mathrm{Fe}}}} x_i\,y_k\,\frac{-1}{|\vec{r}_i-\vec{r}_k|}\, + \sum_{k<l\,\in\,\mathcal{S}_{\mathrm{Fe}}}y_k\,y_l\,\frac{1}{|\vec{r}_k-\vec{r}_l|} \right\} \ .
\label{eq_Ecoul_Li_e}
\end{align}
Eqs.~\eqref{eq_Ecoul_Li_Fe2_Fe3} and \eqref{eq_Ecoul_Li_e} are precisely equivalent, but the two expressions represent alternative interpretations of the combinatorial Coulomb energy. In Eq.~\eqref{eq_Ecoul_Li_Fe2_Fe3}, the terms of the iron sublattice involve distinct \ce{Fe(II)} or \ce{Fe(III)} ions. The respective contributions in Eq.~\eqref{eq_Ecoul_Li_e} are written in terms of negative electronic charges distributed across a fixed \ce{Fe(III)} sublattice (\textit{i.e.}, \ce{Fe(II) \equiv Fe(III) + e^-}), where the binary variables $y_k$ describe whether a given \ce{Fe(III)} sites is carrying an extra redox electron ($y_k = 1$) or not ($y_k = 0$). The second and third terms at the right-hand side of Eq.~\eqref{eq_Ecoul_Li_e} represent the interactions between the \ce{Fe(III)} sublattice and other fixed species (oxygen and phosphor) as well as \ce{Fe(III)}--\ce{Fe(III)} interactions. The linear term in $x_i$ captures the interactions between \ce{Li^+} and fixed species (including the \ce{Fe(III)} sites). The linear term in $y_k$ describes the interactions between the redox \ce{e^-} and fixed species, with the notable exception that for each redox electron, the respective host \ce{Fe(III)} ion is excluded from the summation, as required to avoid divergence of the on-site point-charge interaction. Such divergence is avoided in quantum-mechanical treatments, where the redox electrons are localized in atomic orbitals of the Fe(III) site, yielding a finite on-site interaction energy with the Fe(III) core. Since the latter is identical for all iron sites, adding such an on-site interaction to the energy model of Eq.~\eqref{eq_Ecoul_Li_e} would merely produce a shift in the electronic chemical potential~\cite{saubanereIntuitiveEfficientMethod2014}, but not affect the relative energies of different electronic distributions across the Fe(III) sublattice. In particular, any such on-site contribution to the chemical potential is cancelled by the Legendre transformation within the grand-canonical optimization method employed in this study.

The Coulomb energy model of Eq.~\eqref{eq_Ecoul_Li_e} has the form of a cluster expansion, Eq.~\eqref{eq_cluster_expansion}, which is exact to second order, \textit{i.e.}, all higher-order ECI coefficients are precisely zero. It thus represents a QUBO problem that is suitable for quantum annealing, 
\begin{align}
& E_{\mathrm{coul}}(\{x_i\},\{y_k\}) = E_0 + \sum_{i\,\in\,\mathcal{S}_{\mathrm{Li}}}x_i\,Q_{ii}\ +\ \sum_{k\,\in\,\mathcal{S}_{\mathrm{Fe}}}y_k\,Q_{kk} + \nonumber \\[0.2cm] 
& + \sum_{i<j\,\in\,\mathcal{S}_{\mathrm{Li}}}x_i\,x_j\,Q_{ij}\, + \sum_{\substack{i\,\in\,\mathcal{S}_{\mathrm{Li}}\\ k\,\in\,\mathcal{S}_{\mathrm{Fe}}}} x_i\,y_k\,\,Q_{ik} + \sum_{k<l\,\in\,\mathcal{S}_{\mathrm{Fe}}}y_k\,y_l\,\,Q_{kl} \ .
\label{eq_Ecoul_Li_e_QUBO}
\end{align}
As seen in Eq.~\eqref{eq_Ecoul_Li_e}, the respective QUBO coefficients ($Q_{ij}$, etc.) are given in analytical form as sums over pairwise Coulomb interactions. It should be noted that, as an alternative to the Coulomb model used here, suitable QUBO models can also be obtained from a second-order truncated cluster expansion with ECI coefficients fitted based on DFT energies. For the purpose of the QA method developed in this work, the details of the underlying QUBO model are less important. As shown by the results, however, the Coulomb model is indeed very interesting from a materials science perspective, since it correctly describes the two-phase charging characteristics of LFP and thus captures the essential physics of the process.

\paragraph*{Ewald summation.}
The QUBO coefficients of the \ce{Li^+}--\ce{e^-} model (Eq.~\eqref{eq_Ecoul_Li_e}/\eqref{eq_Ecoul_Li_e_QUBO}) are obtainable by summation of the Coulomb interactions among certain subsets of species/sites. To this end, LFP cells comprising selected subsets of species/sites were constructed and their Coulomb energy evaluated. However, owing to the periodicity of the LFP lattice, each site of the simulation cell, \textit{i.e.}, each of the binary variables, actually represents an entire array of periodic image sites that are simultaneously occupied or vacant. To account for the periodic boundary conditions, the energies of the various sub-structures were calculated using Ewald summation routines of the Python Materials Genomics (\texttt{pymatgen}) library~\cite{ongPythonMaterialsGenomics2013}. The QUBO coefficients $Q_{ij}$, etc., were then obtained after correcting the Ewald energies for certain overcounted interactions. Specifically, the constant energy term, $E_0$, was calculated as the Ewald energy of an LFP cell comprising only fixed species, \textit{i.e.}, \ce{P}, \ce{O}, and with the iron sublattice fully occupied by \ce{Fe(III)}. The linear (diagonal) coefficients $Q_{ii}$ were determined from LFP cells comprising all fixed species (including \ce{Fe(III)}) plus one occupied \ce{Li^+} site ($i$). Because such cells also contained interactions among the fixed species (already captured by $E_0$), the constant term $E_0$ had to be subtracted from the respective Ewald energies to obtain the $Q_{ii}$ coefficients. Similarly, the linear (diagonal) coefficients $Q_{kk}$ were determined from LFP cells comprising all fixed species (including \ce{Fe(III)}) with one iron site ($k$) occupied by \ce{Fe(II)} instead of \ce{Fe(III)}. Again, the respective Ewald energies had to be corrected for the constant term $E_0$ to obtain the $Q_{kk}$. Finally, the quadratic (off-diagonal) coefficients $Q_{ij}$, $Q_{ik}$, and $Q_{kl}$ were determined from LFP cells comprising pairs of two \ce{Li^+} occupying sites $i$ and $j$, one \ce{Li^+} and one \ce{e^-} occupying sites $i$ and $k$, or two \ce{e^-} occupying sites $k$ and $l$, respectively. Here, the respective Ewald energies had to be corrected for the self-interaction of each occupied site with its own periodic images by subtracting the Ewald energies of LFP cells with each single site of a given pair being occupied. 


\subsection*{Configurational sampling}

\paragraph*{Quantum annealing.}
QA was performed on the D-Wave Advantage\texttrademark{} 5.4 System ``JUPSI'' via the \texttt{D-Wave Ocean} interface for Python. The fully connected Coulomb QUBO was mapped to the working graph of the quantum annealer using the \texttt{DWaveCliqueSampler()} routine. An annealing time of $100\,\mathrm{\mu s}$ was used for the individual runs.

\paragraph*{Benchmark sampling/heuristics.}

The QA sampling statistics were analysed by normalization with respect to the joint electro-ionic density of states of the Coulomb model, which was determined by exhaustive sampling (full enumeration) for the $1\times2\times2$ and $1\times4\times1$ supercells. For the analysis of the scaling of computation time with system size, the density of states of the $2\times4\times4$ and $3\times6\times6$ cells were determined by random sampling of ca. 50 million configurations, as implemented in the GOAC package (Global Optimization of Atomistic Configurations by Coulomb)~\cite{kosterOptimizationCoulombEnergies2024}. The Mathematica software~\cite{Mathematica} was employed for fits of the sampled DOS with the log-normal model (Eq.~\eqref{eq_lognorm_DOS}), shown in Fig.~\ref{fig_DOS}a, as well as the numerical calculation of the effective partition function (Eq.~\eqref{eq_partition_function}), shown in Figs.~\ref{fig_DOS}b,c.

The electro-ionic Coulomb model was validated in predicting the known phase-separated ground state of partially lithiated LFP. To this end, the coefficients of the optimization model were calculated with the GOAC package~\cite{kosterOptimizationCoulombEnergies2024} and the model was solved with the Gurobi optimizer~\cite{GurobiOptimizerReference2024} to explore the ground state configuration in semi-lithiated LFP supercells of increasing size, demonstrating the transition to a \ce{LiFePO4}/\ce{FePO4} phase separation within supercells above a critical size in $b$-direction.

\subsection*{Density Functional Theory}

The classical point-charge Coulomb model was compared to DFT simulations accounting for the quantum-mechanical nature of the electronic degrees of freedom. For this purpose, 100 different configurations of Li ions in the semi-lithiated $1\times4\times1$ supercell ($N_{\ce{Li^+}} = 8$ out of 16 total Li sites) were selected by Monte Carlo sampling with the Coulomb model, including the configurations with the absolute highest and lowest Coulomb energies. DFT energies of the selected configurations were calculated at the level of DFT+$U$ using the Quantum Espresso software package~\cite{giannozziQUANTUMESPRESSOModular2009} with ultrasoft pseudopotentials~\cite{vanderbiltSoftSelfconsistentPseudopotentials1990} and the GGA-PBE~\cite{perdewGeneralizedGradientApproximation1996} exchange-correlation functional. The electronic wavefunctions were expanded in a plane-wave basis set up to a cutoff energy of 50\,Ry, and the dispersion of the band structure within the Brillouin zone was sampled with a $3 \times 1 \times 5$ $k$-point mesh. A Hubbard parameter $U = 4.3\,\mathrm{eV}$ was employed for iron $d$-orbitals~\cite{zhouFirstprinciplesPredictionRedox2004}. For 32 out of the 100 configurations, the DFT calculations did not converge, for which reason the Coulomb \textit{vs.} DFT energies of only 68 configurations are compared in Fig.~\ref{fig_LFP_model}b. It should be noted that in previous studies by some of the authors, slightly different values of the Hubbard $U$ parameter were used for Fe(II) and Fe(III) species, which was shown to be essential to accurately describe the formation enthalpies of \ce{FePO4} and \ce{LiFePO4}~\cite{kowalskiElectrodeElectrolyteMaterials2021}. However, these aspects have an effect of about $0.2\,\mathrm{eV}$, which is at least an order of magnitude smaller compared to the energy scales considered in the present work, and thus are neglected here.

\section*{Author Contributions}

TB developed the concept and performed and analysed the quantum annealing calculations. YYT performed the DFT calculations, KK the benchmark optimization and DOS sampling with classical heuristics, and NB the numerical analysis of the effective partition function for computing time estimation. All authors co-wrote, read, and approved the final manuscript.

\section*{Acknowledgements}

The authors gratefully acknowledge the J\"ulich Supercomputing Centre (\url{https://www.fz-juelich.de/ias/jsc}) for funding this project by providing computing time on the D-Wave Advantage\texttrademark{} System JUPSI through the J\"ulich UNified Infrastructure for Quantum computing (JUNIQ) within the project qdisk. DFT simulations were performed on the JURECA machine in the scope of the project cjiek61. P.K. acknowledges the funding from the BMBF under grant number 13XP0558A (AdamBatt2). The presented work was carried out within the framework of the Helmholtz Association’s program Materials and Technologies for the Energy Transition, Topic 2: Electrochemical Energy Storage.

\section*{Competing Interests}

All authors declare no financial or non-financial competing interests.

\section*{Data Availability}

The datasets generated and/or analysed during the current study are not publicly available due to the extensive size of the datasets, but are available from the corresponding author on reasonable request.


\end{document}